\documentclass[reprint,superscriptaddress,longbibliography,amsmath,amssymb,aps,prl]{revtex4-2}

\usepackage{epsfig}
\usepackage{amsfonts}
\usepackage{amsmath}

\begin{document}

\title{Absence of high-field spin supersolid phase in Rb$_2$Co(SeO$_3$)$_2$ with a triangular lattice}

\author{K. Shi}
\thanks{These authors contributed equally to this work.}
\affiliation{Anhui Key Laboratory of Low-Energy Quantum Materials and Devices, High Magnetic Field Laboratory, Chinese Academy of Sciences, Hefei 230031, China}
\affiliation{Hefei National Research Center for Physical Sciences at the Microscale, University of Science and Technology of China, Hefei 230026, China}
\author{Y. Q. Han}
\thanks{These authors contributed equally to this work.}
\affiliation{Anhui Key Laboratory of Low-Energy Quantum Materials and Devices, High Magnetic Field Laboratory, Chinese Academy of Sciences, Hefei 230031, China}
\affiliation{Hefei National Research Center for Physical Sciences at the Microscale, University of Science and Technology of China, Hefei 230026, China}
\author{B. C. Yu}
\affiliation{Key Laboratory of Polar Materials and Devices (MOE), School of Physics and Electronic Science, East China Normal University, Shanghai 200241, China}
\author{L. S. Ling}
\affiliation{Anhui Key Laboratory of Low-Energy Quantum Materials and Devices, High Magnetic Field Laboratory, Chinese Academy of Sciences, Hefei 230031, China}
\author{W. Tong}
\affiliation{Anhui Key Laboratory of Low-Energy Quantum Materials and Devices, High Magnetic Field Laboratory, Chinese Academy of Sciences, Hefei 230031, China}
\author{C. Y. Xi}
\affiliation{Anhui Key Laboratory of Low-Energy Quantum Materials and Devices, High Magnetic Field Laboratory, Chinese Academy of Sciences, Hefei 230031, China}
\author{T. Shang}
\email{tshang@phy.ecnu.edu.cn}
\affiliation{Key Laboratory of Polar Materials and Devices (MOE), School of Physics and Electronic Science, East China Normal University, Shanghai 200241, China}
\affiliation{Chongqing Key Laboratory of Precision Optics, Chongqing Institute of East China Normal University, Chongqing 401120, China}
\author{Zhaosheng Wang}
\email{zswang@hmfl.ac.cn}
\affiliation{Anhui Key Laboratory of Low-Energy Quantum Materials and Devices, High Magnetic Field Laboratory, Chinese Academy of Sciences, Hefei 230031, China}
\author{Li Pi}
\affiliation{Anhui Key Laboratory of Low-Energy Quantum Materials and Devices, High Magnetic Field Laboratory, Chinese Academy of Sciences, Hefei 230031, China}
\affiliation{Hefei National Research Center for Physical Sciences at the Microscale, University of Science and Technology of China, Hefei 230026, China}
\author{Long Ma}
\email{malong@hmfl.ac.cn}
\affiliation{Anhui Key Laboratory of Low-Energy Quantum Materials and Devices, High Magnetic Field Laboratory, Chinese Academy of Sciences, Hefei 230031, China}

\date{\today}

\begin{abstract}
  Magnetization, torque magnetometry, specific heat and nuclear magnetic resonance (NMR) are used to study the high field intermediate phase between
  the 1/3-magnetization plateau and polarized state in the quantum Ising antiferromagnet Rb$_2$Co(SeO$_3$)$_2$ with a triangular lattice. The magnetic phase diagram with the magnetic field up to 30 T is mapped by the comprehensive experimental data. The "up-up-down" (UUD) spin configuration of the 1/3-magnetization plateau state is identified by NMR spectral analysis. At higher magnetic fields, this UUD structure persist to the intermediate phase, which is finally destroyed in the polarized state. This observation supplies unambiguous spectroscopic evidence for the absence of proposed high field spin supersolid phase. The high-field phase diagram of this quantum magnet proximate to the Ising-anisotropy limit contradicts with that proposed by theoretical studies.
\end{abstract}

\maketitle

The unfailing research interest on geometrically frustrated antiferromagnets can historically trace back to the first proposal of the novel resonating valance bond state in spin-1/2 triangular lattice (TL) Heisenberg antiferromagnet in 1970s\cite{Anderson_MRB_1973}. The enhanced quantum fluctuations as result of the failure of simultaneously minimized interaction energies, is believed to strongly renormalize the phases of quantum matter, leading to rich and novel quantum states, quantum phase transitions as well as quantum excitations\cite{Moessner_PhyT_2006}. The quantum phases can be altered by introducing perturbations such as external magnetic field, spin anisotropy, anisotropic bond-dependent couplings, etc.\cite{Bordelon_NP_15_1058,Yunoki_PRB_74_014408,Li_PRB_94_035107,Zhu_PRL_120_207203}. By introducing easy-axis anisotropy to the spin-1/2 TL antiferromagnets, a Y-phase, namely the spin supersolid in the boson language, appears under zero magnetic field. With increasing axial magnetic field intensity, the spin system successively enters the "up-up-down" (UUD) phase, the other supersolid phase (V-phase), and finally the fully polarized state, as indicated by calculations based on spin-1/2 XXZ model on the TL\cite{Yamamoto_PRL_112_127203}.

Recently, two kinds of cobalt-based material Na$_2$BaCo(PO$_4$)$_2$\cite{Zhong_PNAS_116_14505,Li_NC_11_4216,Xiang_Nature_625_270,Sheng_PNAS_119_e2211193119} and K$_2$Co(SeO$_3$)$_2$\cite{Zhong_PRM_4_084406,Zhu_PRL_133_186704,Chen_arxiv_2402_15869} are reported to realize the spin-1/2 XXZ model on TL, owing to the spin-orbit coupling of Co$^{2+}$-ions. In the former material, the Ising and transverse exchange constant, $J_{zz}$ and $J_{\perp}$ are respectively determined to be 1.48 K and 0.88 K, corresponding to a anisotropy ratio $\alpha\equiv J_{\perp}/J_{zz}$ of $\sim0.6$\cite{Xiang_Nature_625_270,Sheng_PNAS_119_e2211193119}. Comparatively, in K$_2$Co(SeO$_3$)$_2$, the $J_{zz}$ and $\alpha$ respectively equal to 34.2 K and 0.07\cite{Zhu_PRL_133_186704,Chen_arxiv_2402_15869}, indicating that the spin system locates much closer to the Ising limit in the phase diagram of spin-1/2 XXZ model\cite{Yamamoto_PRL_112_127203}. The features of solid and superfluid, $\sqrt{3}\times\sqrt{3}$ magnetic unit cell for the $z$-component and gapless Goldstone modes contributed by in-plane spin components, are identified by neutron scattering in both compounds\cite{Xiang_Nature_625_270,Zhu_PRL_133_186704,Chen_arxiv_2402_15869}. These observations provide convincing evidence for the existence of low-field spin supersolid Y-phase. In Na$_2$BaCo(PO$_4$)$_2$, the other supersolid V-phase is further identified for fields above the wide UUD region due to weak Ising-anisotropy\cite{Xiang_Nature_625_270,Gao_NPJ_7_89}. For K$_2$Co(SeO$_3$)$_2$, a similar intermediate region between the UUD and fully polarized phase in the phase diagram is observed by magnetization measurements\cite{Chen_arxiv_2402_15869}. Whether the V-phase exists in the material where the spin anisotropy is much closer to the Ising-limit still lack further spectroscopic studies.

In this paper, we have performed combined experimental study on the high-field intermediate phase in the counterpart material Rb$_2$Co(SeO$_3$)$_2$ using magnetization, torque magnetometry, specific heat and nuclear magnetic resonance (NMR) measurements. The phase diagram with field intensity up to $\mu_0H=30$ T parallel to the $c$-axis and temperature down to $T=1.7$ K is determined from our comprehensive data. The low temperature UUD phase is indicated by the 1/3-magnetization plateau in the field dependence of magnetization and the splitting NMR spectra with a intensity ratio very close to 2:1. Under stronger magnetic field, the subtle intermediate phase manifests itself via the hump behavior in the field dependence of magnetization, magnetic torque and specific heat. From NMR as a microscopic probe, both central peaks respectively contributed by the up- and down-spin sublattice simultaneously shift to a slightly lower frequency relative to the Larmor frequency when the spin system enters the intermediate phase from the UUD-phase. The field dependence of spin-lattice relaxation rates shows clear slope change behavior across the field induced transition. However, the frequency difference between the splitting central peaks remains unchanged across the transition, until the spin system is polarized. This observation rules out the possibility of spin-reorientation across the transition, thus supply compelling evidence for the absence of high-field spin supersolid phase (V-phase) in Rb$_2$Co(SeO$_3$)$_2$.

Single crystals of Rb$_2$Co(SeO$_3$)$_2$ were synthesized using a solid-state reaction method as described elsewhere\cite{Zhong_PRM_4_084406}. The high magnetic field was generated by water-cooled resistive magnets at CHMFL. We performed $dc$-magnetization and magnetic torque measurements respectively using vibrating sample magnetometer and the cantilever-based magnetometry with a capacitive readout. Specific heat were measured by a long relaxation method\cite{Wang_PhysicaC_355_179,Taylor_PRL_99_057001}, where the thermometer was a calibrated Cernox chip resistor. The NMR measurements were performed on $^{87}$Rb nuclei ($\gamma_n=13.931$ MHz/T, $I=3/2$) with a phase-coherent NMR spectrometer. The spectrum was obtained by integrating the spectral intensity with the frequency swept. The spin-lattice relaxation rate ($1/T_1$) was measured with the inversion-recovery sequence.

\begin{figure}
  \includegraphics[width=9cm, height=11cm]{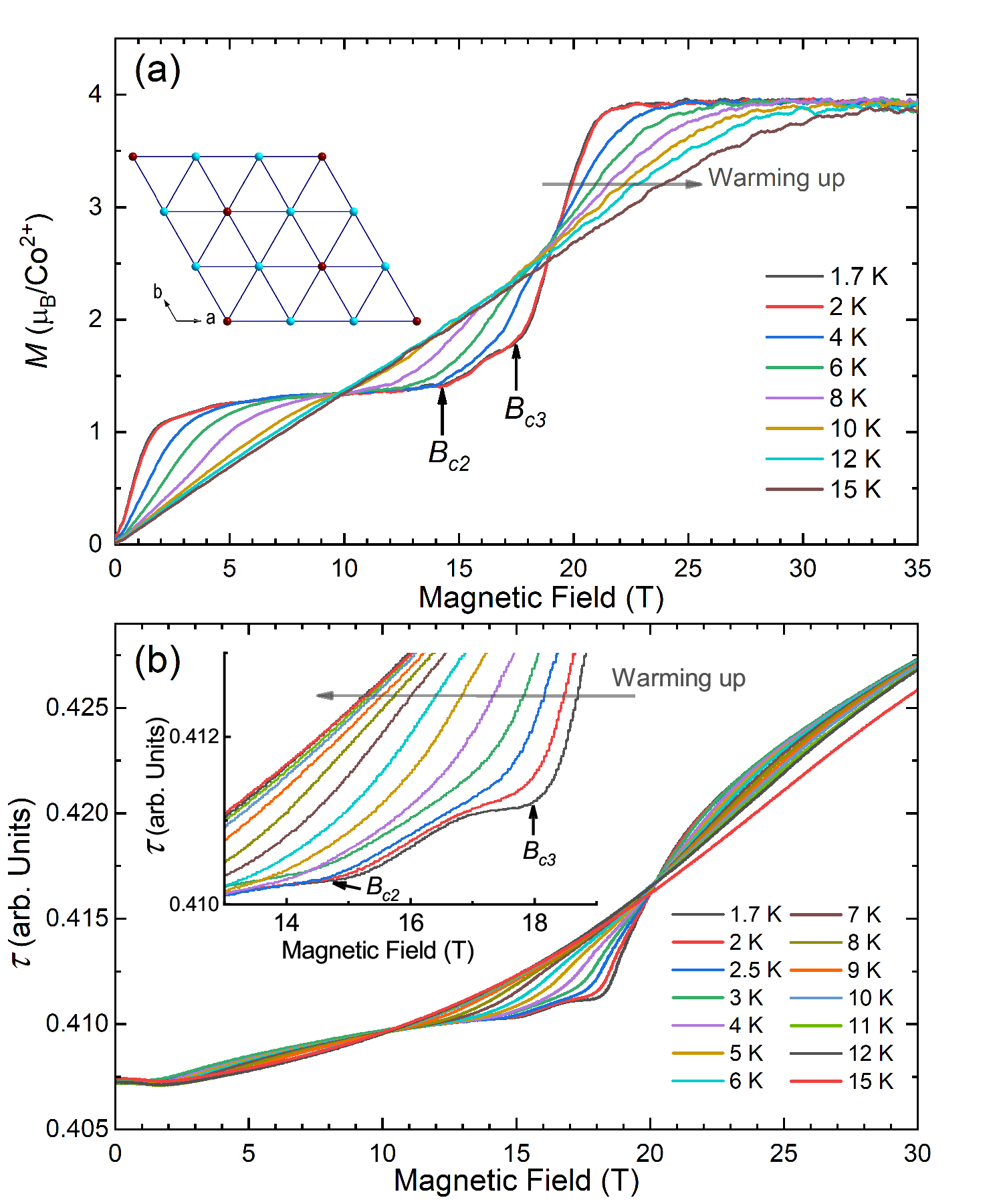}
  \caption{\label{mag1}(color online) The $c$-axis magnetic field dependence of $dc$-magnetization (a) and magnetic torque (b) at temperatures ranging from 1.7 K to 15 K. The UUD spin configuration in the 1/3-$M_s$ plateau state is shown in the inset of (a) as observed against $c$-axis. The cyan and red balls respectively denote the magnetic sites with moments pointing upwards and downwards. In the inset of (b), a enlarge view of the field-dependent magnetic torque is shown to demonstrate the high-field intermediate phase.
}
\end{figure}

We begin from identifying the magnetic phase diagram with macroscopic probes. In Fig.\ref{mag1}(a), the low temperature $dc$-magnetization is shown as a function of magnetic field intensity. At the first glance, the low temperature ($T=1.7$ K or 2 K) magnetization first steeply increases, then enters a plateau region for fields between $\mu_0H\sim 2$ T and $\sim14$ T, above which the magnetization again increases and finally reaches a saturation value of $M_s\sim3.9 \mu_B/$Co$^{2+}$ above $\mu_0H\sim 21$ T. This is similar with that observed in K$_2$Co(SeO$_3$)$_2$\cite{Zhu_PRL_133_186704}. More importantly, the magnetization at fields between $\mu_0H\sim 14.3$ T and $\sim17.4$ T shows up a clear hump behavior, differing from that in the other frustrated antiferromagnets\cite{Susuki_PRL_110_267201,Xing_PRB_100_220407,Ono_PRB_67_104431}. The $J_{zz}$ and $\alpha$-value are very close to those in the counterpart material K$_2$Co(SeO$_3$)$_2$, as determined from the characteristic field intensities. Upon warming, the magnetization plateau becomes narrower, the hump behavior becomes undistinguishable above $T=6$ K, and the saturation field increases. For temperatures exceeding $T=10$ K, the magnetization behavior becomes featureless, signaling the breakdown of magnetic order by thermal fluctuations. In Fig.\ref{mag1}(b), the field dependent magnetic torque is presented for the same temperature range. The low temperature hump behavior also turns up in the torque measurement, but in a more pronounced manner (See the inset).

\begin{figure}
\includegraphics[width=9cm, height=5.9cm]{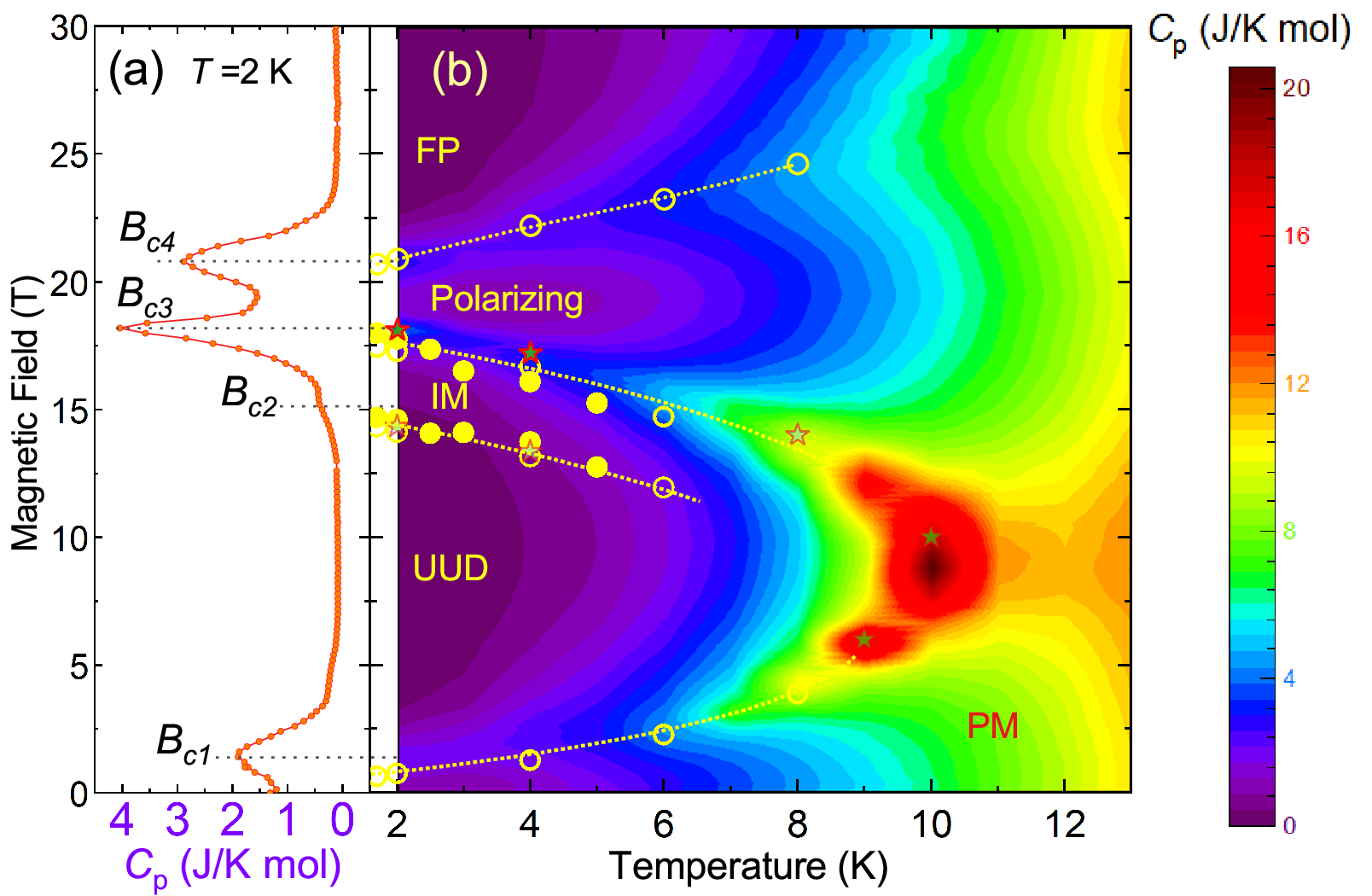}
\caption{\label{pd2}(color online)
  (a) Field dependence of the specific heat at the temperature $T=2$ K. Four characteristic fields are marked by the dotted lines.
  (b) Field-temperature phase diagram of the spin system in Rb$_2$Co(SeO$_3$)$_2$. The background color map is a contour plot of the specific heat versus field and temperature. The solid circles, hollow circles and stars denotes the phase boundary determined from magnetic torque, $dc$-magnetization and NMR measurements.
}
\end{figure}

To get more insights into series of the field-induced phase transitions, we have performed specific heat ($C_p$) measurements on Rb$_2$Co(SeO$_3$)$_2$ with fields up to 30 T, and conduct a contour plot of the data as shown in Fig.\ref{pd2}(b). The characteristic field intensities $B_{c1}$ to $B_{c4}$ are also plotted for precisely determining the phase boundary. The field dependence of $C_p$ at $T=2$ K is shown in Fig.\ref{pd2}(a).
With the magnetic field increasing from zero to 30 T, the $C_p(H)$ exhibits three well-defined peaks and one tiny hump. Thus, the low-temperature phase is divided into five phases,including the paramagnetic, UUD, intermediate, polarizing and fully-polarized phase with the increasing field intensity.

Based on the macroscopic probes, several phases can be determined with further guidance from the theory and previous studies on K$_2$Co(SeO$_3$)$_2$\cite{Yamamoto_PRL_112_127203,Chen_arxiv_2402_15869,Zhu_PRL_133_186704}. As proposed by the theoretical studies on spin-1/2 triangular antiferromagnets\cite{Yamamoto_PRL_112_127203}, the collinear UUD phase is selected by the quantum spin fluctuations, when an external field is applied either in easy-plane or along easy-axis. The phase between $B_{c1}$ and $B_{c2}$ is the $1/3-M_s$ plateau state, precisely consistent with the UUD phase. The phase below $B_{c1}$ is the paramagnetic phase for the present $T\geq1.7$ K, as indicated by previous studies\cite{Chen_arxiv_2402_15869,Zhu_PRL_133_186704}. A spin supersolid phase (Y-phase) appears at much lower temperatures\cite{Chen_arxiv_2402_15869,Zhu_PRL_133_186704}. For fields above $B_{c3}$, the $dc$-magnetization shows a steep increase, and finally saturate for fields above $B_{c4}$. Thus, these two regions correspond to the polarizing and fully-polarized phase.
The intermediate phase locating between $B_{c2}$ and $B_{c3}$ is previously proposed to be the second spin supersolid phase (V-phase) based on the $M(T)$ measurements under high fields\cite{Chen_arxiv_2402_15869}. This conclusion is naturally drawn by the theoretical predictions as well as the realization of high field spin supersolid phase in another triangular antiferromagnet Na$_2$BaCo(PO$_4$)$_2$\cite{Xiang_Nature_625_270,Gao_NPJ_7_89}. However, this conjecture still lack further spectroscopic evidence.

\begin{figure}
\includegraphics[width=9cm, height=10.8cm]{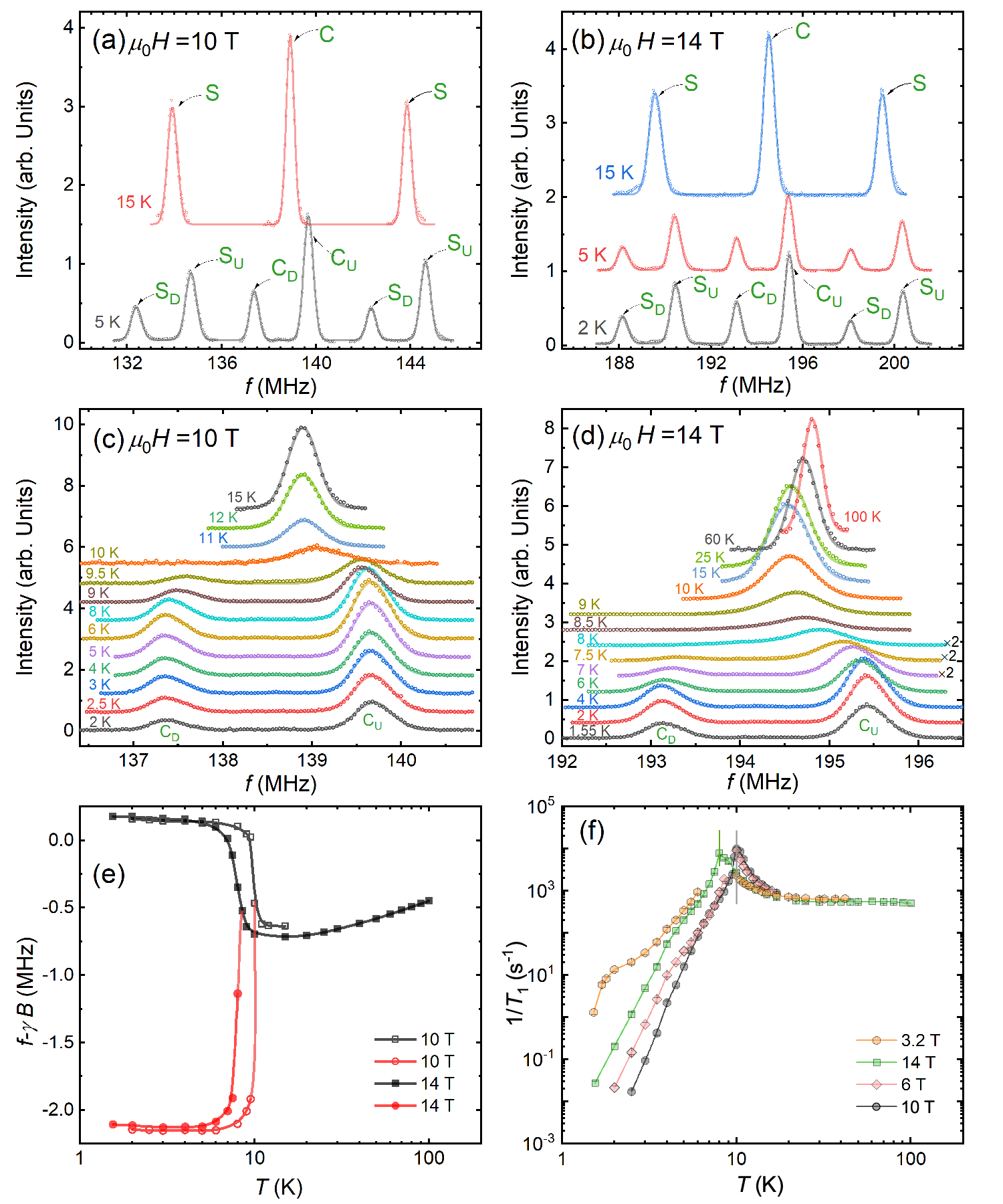}
\caption{\label{nmr3}(color online)
  Full $^{87}$Rb spectra at the paramagnetic and magnetically ordered states under fields of 10 T (a) and 14 T (b). C and S respectively represent central and satellite peaks. The subscripts U and D label the peak contributed by sublattice with magnetic moments parallel or antiparallel to the applied field. (c)and (d): Temperature dependence of the central peak at magnetic fields of 10 T and 14 T. The solid lines in (a)-(d) are fits to single- or Multi-Gaussian peak functions. (e): Resonance frequency as a function of temperature. (f) The temperature dependence of spin-lattice relaxation rates for field ranged from 3.2 T to 14 T.
}
\end{figure}

We employ NMR as a microscopic probe to study the spin configuration under different magnetic fields. Full $^{87}$Rb NMR spectra under a magnetic field of 10 T and 14 T are respectively shown in Fig.\ref{nmr3}(a) and (b). Based on the phase diagram shown in Fig.\ref{pd2}, the spin system enters the UUD phase below $T_N=10$ K under the 10 T field. While, for $\mu_0H=14$ T, double phase transitions occur upon cooling, including the paramagnetic to intermediate phase at $T_{N1}=8$ K, and the intermediate to the UUD phase at $T_{N2}=3$ K. At $T=15$ K, three resonance peaks including single central peak and double satellite peaks symmetrically locating at both sides, are detected under both magnetic fields. This is consistent with the single Wyckoff position of $^{87}$Rb carrying nuclear spin $I=3/2$\cite{Zhong_PRM_4_084406}. Upon cooling, each peak splits into two peaks with a relative spectral weight close to 2:1, signaling the setup of magnetic order. Surprisingly, no noticeable difference is observed between the spectrum for the intermediate phase and UUD phase, as shown in Fig.\ref{nmr3}(b).

The temperature dependent central peak under the field of 10 T and 14 T are further shown in Fig.\ref{nmr3}(c) and (d) to track the phase evolution. The resonance frequency are obtained by fitting the peak with the gaussian function. Fig.\ref{nmr3}(e) shows the central peak frequency shift relative to the Larmor frequency as a function of temperature. In the PM phase, the central line with a negative Knight shift, shifts to the same low-frequency-side with the enhanced uniform spin susceptibility upon cooling. This indicates a negative diagonal element $A_{cc}$ of the hyperfine coupling matrix $\{A_{ij}\}$. The critical temperature identified as $T_N=10$ K for the applied 10 T field, and 8 K for the 14 T field, below which the NMR line splits. No abrupt frequency change occurs when the spin system cross the phase boundary between the UUD and intermediate phase at $T_{N2}$. We have also measured the spin-lattice relaxation rates at the central peak frequency, whose temperature dependence under different magnetic fields are shown in Fig.\ref{nmr3}(f). The magnetic phase transitions manifest themselves via the "$\lambda$"-peaks centered at $T_N$ or $T_{N1}$ contributed by the critical slowing down of the spin system.

\begin{figure*}
\includegraphics[width=15cm, height=4.8cm]{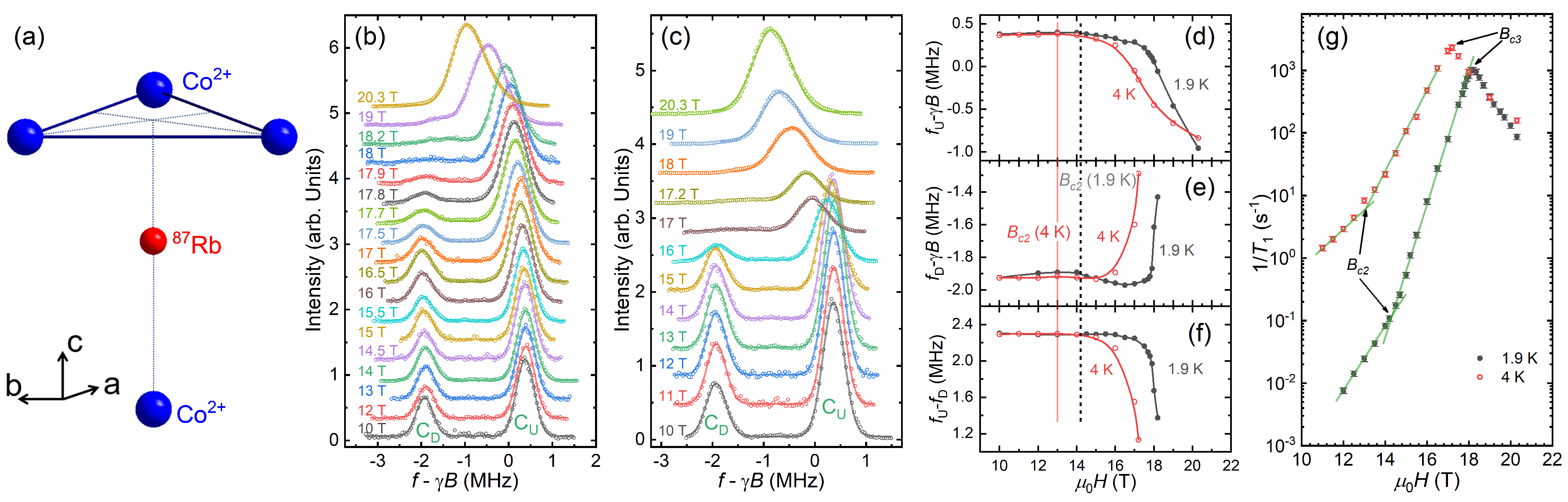}
\caption{\label{nmr4}(color online)
  (a) A sketch of the studied $^{87}$Rb nuclear sites and the neighboring Co$^{2+}$ magnetic sites.
  (b) and (c) Central peak at $T=1.9$ K and $T=4$ K under different magnetic fields. Solid lines are fits to single- or multi-Gaussian peak function.
  (d)-(f) Resonance frequencies and the frequency difference between splitting peaks as a function of field intensity.
  (g) Field dependence of the spin-lattice relaxation rates at $T=1.9$ K and $T=4$ K.
}
\end{figure*}

The splitting NMR lines at low temperatures can be understood with the UUD spin configuration. In Fig.\ref{nmr4}(a), we present a schematic description of the local environment of $^{87}$Rb nuclei with respect to the magnetic Co$^{2+}$ sites. In Rb$_2$Co(SeO$_3$)$_2$, layers of Co$^{2+}$-triangles are stacked with an ABC-type periodicity in the crystalline $c$-axis\cite{Zhong_PRM_4_084406}. The Rb$^{+}$-ions occupy the interstitial position, locating above/below the center of Co$^{2+}$-triangles, and below/above Co$^{2+}$-ions in the adjacent Co$^{2+}$-layer along the $c$-axis. For the UUD magnetic structure, the internal hyperfine field must be aligned with the crystalline $c$-axis as constrained by the local symmetry. The Co$^{2+}$-triangles contribute the same internal field, while the Co$^{2+}$-ions carrying upward or downward magnetic moments from the adjacent layers contribute staggered internal field.
The intensity ratio of splitting resonance peaks can be intuitively expected to be 2:1, as the amount of Co$^{2+}$-ions carrying upward moments is two times of that carrying downward moments\cite{Sakhratov_PRB_99_024419}. Change of the moment direction or size will further modify the frequency difference between the splitting lines. Thus, our observations in the UUD phase is consistent with the field induced magnetic structure with a quantum origin. We propose that in the intermediate phase, the spin system also adopt the same UUD spin configuration.

To eliminate the influence of thermal fluctuations, we have studied the evolution of NMR spectra and spin-lattice relaxation rates at constant temperatures with the applied magnetic field, whose range covers the UUD, intermediate and polarized phase. The central lines at temperatures of 1.9 K and 4 K are respectively shown in Fig.\ref{nmr4}(b) and (c), with the field intensity varying from 10 T to 20.3 T. At $T=1.9$ K, well-defined double central lines are observed for the field ranged from 10 T to $\sim 17.9$ T, above which the intensity of the low-frequency-side peak is quickly suppressed. Single central peak is observed at $\mu_0H=20.3$ T, indicating the emergence of spin polarized state. For $T=4$ K, one observes similar behaviors, just at relatively lower characteristic field intensities.

In Fig.\ref{nmr4}(d)-(f), the frequency of each resonance peak, and differences between them are plotted as a function of field intensity for temperatures of 1.9 K and 4 K. The field dependence of the spin-lattice relaxation rates at $T=1.9$ K and 4 K is shown in Fig.\ref{nmr4}(g). The critical fields $B_{c2}$ and $B_{c3}$ respectively manifest themselves via the abrupt change of the slope and a peaked behavior in the field dependent $1/T_1$. The $B_{c2}$'s for $T=2$ K and 4 K are also marked with the vertical lines in Fig.\ref{nmr4}(d)-(f). Clearly, frequency difference between two peaks does not change across the phase boundary. Instead, both peaks shift toward low-frequency-side. As a result of the negative diagonal element $A_{cc}$ of the hyperfine coupling matrix, this uniform line shift is consistent with the increasing bulk magnetization above $B_{c1}$ as shown in Fig.\ref{mag1}(a).

Our observations directly indicate the persistent UUD spin configuration in the intermediate phase. When the spin system approaching the polarized state, the intensity of central line marked by C$_D$ is strongly suppressed, and the peak shifts to higher frequencies. This indicates a spin reorientation under high magnetic fields. The strongly suppressed intensity precisely reflects that spins pointing up or down is energetically favored instead of other directions, again demonstrating the Ising anisotropy in the present sample. Contrastingly, the C$_U$-line slightly shifts to the low-frequency-side across the $B_{c2}$. No obvious change is observed at the relative intensity of C$_D$-peak at the meanwhile. The staggered hyperfine field measured by the frequency difference between the C$_U$ and C$_D$ peak stays as a constant for the UUD phase and intermediate phase. This is completely different with what occurs in Na$_2$BaCo(PO$_4$)$_2$\cite{Xu_arxiv_2504_08570}. Our data supply unambiguous evidence for the absence of the high-field spin supersolid phase in Rb$_2$Co(SeO$_3$)$_2$.

The magnetic phase diagram under high magnetic fields obviously deviates from that proposed by theoretical studies. Based on the triangular-lattice spin-1/2 XXZ model, the cluster mean-field study provides a complete quantum phase diagram\cite{Yamamoto_PRL_112_127203}. For the present sample close to the Ising-anisotropy limit,
one expects a high field V-phase between the UUD and polarized phase. However, a very recent numerical study finds that the spin system will adopt the $\pi$-coplanar configuration for bond dimensions larger than 1\cite{Xu_PRB_111_L060402}. From the spin dynamics, continuum of spin excitations are observed in the spin supersolid phase of both Na$_2$BaCo(PO$_4$)$_2$ and K$_2$Co(SeO$_3$)$_2$ compounds\cite{Sheng_Innovation_6_100769,Zhu_PRL_133_186704}. Compared with the spin-wave description of the spin excitation spectrum, the spinon language is proposed to supply a more precise starting point\cite{Zhu_PRL_133_186704,Jia_PRR_6_033031}. The spin system may be proximate to the quantum spin liquid. Compared with Na$_2$BaCo(PO$_4$)$_2$\cite{Xiang_Nature_625_270,Sheng_PNAS_119_e2211193119}, the spin system in Rb$_2$Co(SeO$_3$)$_2$ is much closer to the Ising-anisotropy limit in the phase diagram\cite{Zhu_PRL_133_186704,Chen_arxiv_2402_15869}. Strong quantum fluctuations may prominently renormalize the quantum ground state under high magnetic field, and lead to the absence of high field spin supersolid phase in Rb$_2$Co(SeO$_3$)$_2$. Further theoretical studies starting from spinon representations are needed to verify this speculation.

To summarize, the high-field intermediate phase between the UUD and polarized phase in Rb$_2$Co(SeO$_3$)$_2$ is investigated by comprehensive experimental methods, including $dc$-magnetization, torque magnetometry, specific heat and NMR. The magnetic phase diagram is established with field intensity up to 30 T. The UUD spin configuration is identified by the NMR spectral analysis in the 1/3-magnetization plateau state. When the spin system enters the intermediate phase, the frequency difference and intensity ratio between the splitting NMR peaks stay unchanged. Finally, the spectral weight contributed by sublattices with spins pointing down is suppressed to zero for the polarized phase. These observations unambiguously indicate the persistent UUD structure in the intermediate phase, thus the absence of the high-field spin supersolid phase in Rb$_2$Co(SeO$_3$)$_2$. We speculate that the ground state of the quantum Ising magnet may be strongly renormalized by the strong quantum fluctuations.

This research was supported by the National Natural Science Foundation of China (Grants No. 11874057, 12374105 and 11874359), the Natural Science Foundation of Shanghai (Grants No. 21ZR1420500 and 21JC1402300), Natural Science Foundation of Chongqing (Grant No. CSTB-2022NSCQ-MSX1678),the Basic Research Program of the Chinese Academy of Sciences Based on Major Scientific Infrastructures (Grant No. JZHKYPT-2021-08) and Fundamental Research Funds for the Central Universities. A portion of this work was supported by the High Magnetic Field Laboratory of Anhui Province. We thank the WM2 (https://cstr.cn/31125.02.SHMFF.WM2) at the Steady High Magnetic Field Facility, CAS (https://cstr.cn/31125.02.SHMFF), for providing technical support and assistance in data collection and analysis.

%\bibliography{RCSO_ref}

\begin{thebibliography}{25}%
\makeatletter
\providecommand \@ifxundefined [1]{%
 \@ifx{#1\undefined}
}%
\providecommand \@ifnum [1]{%
 \ifnum #1\expandafter \@firstoftwo
 \else \expandafter \@secondoftwo
 \fi
}%
\providecommand \@ifx [1]{%
 \ifx #1\expandafter \@firstoftwo
 \else \expandafter \@secondoftwo
 \fi
}%
\providecommand \natexlab [1]{#1}%
\providecommand \enquote  [1]{``#1''}%
\providecommand \bibnamefont  [1]{#1}%
\providecommand \bibfnamefont [1]{#1}%
\providecommand \citenamefont [1]{#1}%
\providecommand \href@noop [0]{\@secondoftwo}%
\providecommand \href [0]{\begingroup \@sanitize@url \@href}%
\providecommand \@href[1]{\@@startlink{#1}\@@href}%
\providecommand \@@href[1]{\endgroup#1\@@endlink}%
\providecommand \@sanitize@url [0]{\catcode `\\12\catcode `\$12\catcode
  `\&12\catcode `\#12\catcode `\^12\catcode `\_12\catcode `\%12\relax}%
\providecommand \@@startlink[1]{}%
\providecommand \@@endlink[0]{}%
\providecommand \url  [0]{\begingroup\@sanitize@url \@url }%
\providecommand \@url [1]{\endgroup\@href {#1}{\urlprefix }}%
\providecommand \urlprefix  [0]{URL }%
\providecommand \Eprint [0]{\href }%
\providecommand \doibase [0]{https://doi.org/}%
\providecommand \selectlanguage [0]{\@gobble}%
\providecommand \bibinfo  [0]{\@secondoftwo}%
\providecommand \bibfield  [0]{\@secondoftwo}%
\providecommand \translation [1]{[#1]}%
\providecommand \BibitemOpen [0]{}%
\providecommand \bibitemStop [0]{}%
\providecommand \bibitemNoStop [0]{.\EOS\space}%
\providecommand \EOS [0]{\spacefactor3000\relax}%
\providecommand \BibitemShut  [1]{\csname bibitem#1\endcsname}%
\let\auto@bib@innerbib\@empty
%</preamble>
\bibitem [{\citenamefont {Anderson}(1973)}]{Anderson_MRB_1973}%
  \BibitemOpen
  \bibfield  {author} {\bibinfo {author} {\bibfnamefont {P.}~\bibnamefont
  {Anderson}},\ }\bibfield  {title} {\bibinfo {title} {Resonating valence
  bonds: a new kind of insulator?},\ }\href@noop {} {\bibfield  {journal}
  {\bibinfo  {journal} {Mater. Res. Bull.}\ }\textbf {\bibinfo {volume} {8}},\
  \bibinfo {pages} {153} (\bibinfo {year} {1973})}\BibitemShut {NoStop}%
\bibitem [{\citenamefont {Moessner}\ and\ \citenamefont
  {Ramirez}(2006)}]{Moessner_PhyT_2006}%
  \BibitemOpen
  \bibfield  {author} {\bibinfo {author} {\bibfnamefont {R.}~\bibnamefont
  {Moessner}}\ and\ \bibinfo {author} {\bibfnamefont {A.~P.}\ \bibnamefont
  {Ramirez}},\ }\bibfield  {title} {\bibinfo {title} {Geometrical
  frustration},\ }\href@noop {} {\bibfield  {journal} {\bibinfo  {journal}
  {Phys. Today}\ }\textbf {\bibinfo {volume} {59}},\ \bibinfo {pages} {24}
  (\bibinfo {year} {2006})}\BibitemShut {NoStop}%
\bibitem [{\citenamefont {Bordelon}\ \emph {et~al.}(2019)\citenamefont
  {Bordelon}, \citenamefont {Kenney}, \citenamefont {Liu}, \citenamefont
  {Hogan}, \citenamefont {Posthuma}, \citenamefont {Kavand}, \citenamefont
  {Lyu}, \citenamefont {Sherwin}, \citenamefont {Butch}, \citenamefont {Brown},
  \citenamefont {Graf}, \citenamefont {Balents},\ and\ \citenamefont
  {Wilson}}]{Bordelon_NP_15_1058}%
  \BibitemOpen
  \bibfield  {author} {\bibinfo {author} {\bibfnamefont {M.~M.}\ \bibnamefont
  {Bordelon}}, \bibinfo {author} {\bibfnamefont {E.}~\bibnamefont {Kenney}},
  \bibinfo {author} {\bibfnamefont {C.}~\bibnamefont {Liu}}, \bibinfo {author}
  {\bibfnamefont {T.}~\bibnamefont {Hogan}}, \bibinfo {author} {\bibfnamefont
  {L.}~\bibnamefont {Posthuma}}, \bibinfo {author} {\bibfnamefont
  {M.}~\bibnamefont {Kavand}}, \bibinfo {author} {\bibfnamefont
  {Y.}~\bibnamefont {Lyu}}, \bibinfo {author} {\bibfnamefont {M.}~\bibnamefont
  {Sherwin}}, \bibinfo {author} {\bibfnamefont {N.~P.}\ \bibnamefont {Butch}},
  \bibinfo {author} {\bibfnamefont {C.}~\bibnamefont {Brown}}, \bibinfo
  {author} {\bibfnamefont {M.~J.}\ \bibnamefont {Graf}}, \bibinfo {author}
  {\bibfnamefont {L.}~\bibnamefont {Balents}},\ and\ \bibinfo {author}
  {\bibfnamefont {S.~D.}\ \bibnamefont {Wilson}},\ }\bibfield  {title}
  {\bibinfo {title} {Field-tunable quantum disordered ground state in the
  triangular-lattice antiferromagnet {N}a{Y}b{O}$_2$},\ }\href@noop {}
  {\bibfield  {journal} {\bibinfo  {journal} {Nat. Phys.}\ }\textbf {\bibinfo
  {volume} {15}},\ \bibinfo {pages} {1058} (\bibinfo {year}
  {2019})}\BibitemShut {NoStop}%
\bibitem [{\citenamefont {Yunoki}\ and\ \citenamefont
  {Sorella}(2006)}]{Yunoki_PRB_74_014408}%
  \BibitemOpen
  \bibfield  {author} {\bibinfo {author} {\bibfnamefont {S.}~\bibnamefont
  {Yunoki}}\ and\ \bibinfo {author} {\bibfnamefont {S.}~\bibnamefont
  {Sorella}},\ }\bibfield  {title} {\bibinfo {title} {Two spin liquid phases in
  the spatially anisotropic triangular {H}eisenberg model},\ }\href@noop {}
  {\bibfield  {journal} {\bibinfo  {journal} {Phys. Rev. B}\ }\textbf {\bibinfo
  {volume} {74}},\ \bibinfo {pages} {014408} (\bibinfo {year}
  {2006})}\BibitemShut {NoStop}%
\bibitem [{\citenamefont {Li}\ \emph {et~al.}(2016)\citenamefont {Li},
  \citenamefont {Wang},\ and\ \citenamefont {Chen}}]{Li_PRB_94_035107}%
  \BibitemOpen
  \bibfield  {author} {\bibinfo {author} {\bibfnamefont {Y.~D.}\ \bibnamefont
  {Li}}, \bibinfo {author} {\bibfnamefont {X.~Q.}\ \bibnamefont {Wang}},\ and\
  \bibinfo {author} {\bibfnamefont {G.}~\bibnamefont {Chen}},\ }\bibfield
  {title} {\bibinfo {title} {Anisotropic spin model of strong
  spin-orbit-coupled triangular antiferromagnets},\ }\href@noop {} {\bibfield
  {journal} {\bibinfo  {journal} {Phys. Rev. B}\ }\textbf {\bibinfo {volume}
  {94}},\ \bibinfo {pages} {035107} (\bibinfo {year} {2016})}\BibitemShut
  {NoStop}%
\bibitem [{\citenamefont {Zhu}\ \emph {et~al.}(2018)\citenamefont {Zhu},
  \citenamefont {Maksimov}, \citenamefont {White},\ and\ \citenamefont
  {Chernyshev}}]{Zhu_PRL_120_207203}%
  \BibitemOpen
  \bibfield  {author} {\bibinfo {author} {\bibfnamefont {Z.~Y.}\ \bibnamefont
  {Zhu}}, \bibinfo {author} {\bibfnamefont {P.~A.}\ \bibnamefont {Maksimov}},
  \bibinfo {author} {\bibfnamefont {S.~R.}\ \bibnamefont {White}},\ and\
  \bibinfo {author} {\bibfnamefont {A.~L.}\ \bibnamefont {Chernyshev}},\
  }\bibfield  {title} {\bibinfo {title} {Topography of spin liquids on a
  triangular lattice},\ }\href@noop {} {\bibfield  {journal} {\bibinfo
  {journal} {Phys. Rev. Lett.}\ }\textbf {\bibinfo {volume} {120}},\ \bibinfo
  {pages} {207203} (\bibinfo {year} {2018})}\BibitemShut {NoStop}%
\bibitem [{\citenamefont {Yamamoto}\ \emph {et~al.}(2014)\citenamefont
  {Yamamoto}, \citenamefont {Marmorini},\ and\ \citenamefont
  {Danshita}}]{Yamamoto_PRL_112_127203}%
  \BibitemOpen
  \bibfield  {author} {\bibinfo {author} {\bibfnamefont {D.}~\bibnamefont
  {Yamamoto}}, \bibinfo {author} {\bibfnamefont {G.}~\bibnamefont
  {Marmorini}},\ and\ \bibinfo {author} {\bibfnamefont {I.}~\bibnamefont
  {Danshita}},\ }\bibfield  {title} {\bibinfo {title} {Quantum phase diagram of
  the triangular-lattice ${XXZ}$ model in a magnetic field},\ }\href@noop {}
  {\bibfield  {journal} {\bibinfo  {journal} {Phys. Rev. Lett.}\ }\textbf
  {\bibinfo {volume} {112}},\ \bibinfo {pages} {127203} (\bibinfo {year}
  {2014})}\BibitemShut {NoStop}%
\bibitem [{\citenamefont {Zhong}\ \emph {et~al.}(2019)\citenamefont {Zhong},
  \citenamefont {Guo}, \citenamefont {Xu}, \citenamefont {Xu},\ and\
  \citenamefont {Cava}}]{Zhong_PNAS_116_14505}%
  \BibitemOpen
  \bibfield  {author} {\bibinfo {author} {\bibfnamefont {R.~D.}\ \bibnamefont
  {Zhong}}, \bibinfo {author} {\bibfnamefont {S.}~\bibnamefont {Guo}}, \bibinfo
  {author} {\bibfnamefont {G.~Y.}\ \bibnamefont {Xu}}, \bibinfo {author}
  {\bibfnamefont {Z.~J.}\ \bibnamefont {Xu}},\ and\ \bibinfo {author}
  {\bibfnamefont {R.~J.}\ \bibnamefont {Cava}},\ }\bibfield  {title} {\bibinfo
  {title} {Strong quantum fluctuations in a quantum spin liquid candidate with
  a {C}o-based triangular lattice},\ }\href@noop {} {\bibfield  {journal}
  {\bibinfo  {journal} {Proc. Natl. Acad. Sci.}\ }\textbf {\bibinfo {volume}
  {116}},\ \bibinfo {pages} {14505} (\bibinfo {year} {2019})}\BibitemShut
  {NoStop}%
\bibitem [{\citenamefont {Li}\ \emph {et~al.}(2020)\citenamefont {Li},
  \citenamefont {Huang}, \citenamefont {Yue}, \citenamefont {Chu},
  \citenamefont {Chen}, \citenamefont {Choi}, \citenamefont {Zhao},
  \citenamefont {Zhou},\ and\ \citenamefont {Sun}}]{Li_NC_11_4216}%
  \BibitemOpen
  \bibfield  {author} {\bibinfo {author} {\bibfnamefont {N.}~\bibnamefont
  {Li}}, \bibinfo {author} {\bibfnamefont {Q.}~\bibnamefont {Huang}}, \bibinfo
  {author} {\bibfnamefont {X.~Y.}\ \bibnamefont {Yue}}, \bibinfo {author}
  {\bibfnamefont {W.~J.}\ \bibnamefont {Chu}}, \bibinfo {author} {\bibfnamefont
  {Q.}~\bibnamefont {Chen}}, \bibinfo {author} {\bibfnamefont {E.~S.}\
  \bibnamefont {Choi}}, \bibinfo {author} {\bibfnamefont {X.}~\bibnamefont
  {Zhao}}, \bibinfo {author} {\bibfnamefont {H.~D.}\ \bibnamefont {Zhou}},\
  and\ \bibinfo {author} {\bibfnamefont {X.~F.}\ \bibnamefont {Sun}},\
  }\bibfield  {title} {\bibinfo {title} {Possible itinerant excitations and
  quantum spin state transitions in the effective spin-1/2 triangular-lattice
  antiferromagnet {N}a$_2${B}a{C}o({PO}$_4$)$_2$},\ }\href@noop {} {\bibfield
  {journal} {\bibinfo  {journal} {Nat. Comm.}\ }\textbf {\bibinfo {volume}
  {11}},\ \bibinfo {pages} {4216} (\bibinfo {year} {2020})}\BibitemShut
  {NoStop}%
\bibitem [{\citenamefont {Xiang}\ \emph {et~al.}(2024)\citenamefont {Xiang},
  \citenamefont {Zhang}, \citenamefont {Gao}, \citenamefont {Schmidt},
  \citenamefont {Schmalzl}, \citenamefont {Wang}, \citenamefont {Li},
  \citenamefont {Xi}, \citenamefont {Liu}, \citenamefont {Jin}, \citenamefont
  {Li}, \citenamefont {Shen}, \citenamefont {Chen}, \citenamefont {Qi},
  \citenamefont {Wan}, \citenamefont {Jin}, \citenamefont {Li}, \citenamefont
  {Sun},\ and\ \citenamefont {Su}}]{Xiang_Nature_625_270}%
  \BibitemOpen
  \bibfield  {author} {\bibinfo {author} {\bibfnamefont {J.~S.}\ \bibnamefont
  {Xiang}}, \bibinfo {author} {\bibfnamefont {C.~D.}\ \bibnamefont {Zhang}},
  \bibinfo {author} {\bibfnamefont {Y.}~\bibnamefont {Gao}}, \bibinfo {author}
  {\bibfnamefont {W.}~\bibnamefont {Schmidt}}, \bibinfo {author} {\bibfnamefont
  {K.}~\bibnamefont {Schmalzl}}, \bibinfo {author} {\bibfnamefont {C.~W.}\
  \bibnamefont {Wang}}, \bibinfo {author} {\bibfnamefont {B.}~\bibnamefont
  {Li}}, \bibinfo {author} {\bibfnamefont {N.}~\bibnamefont {Xi}}, \bibinfo
  {author} {\bibfnamefont {X.~Y.}\ \bibnamefont {Liu}}, \bibinfo {author}
  {\bibfnamefont {H.}~\bibnamefont {Jin}}, \bibinfo {author} {\bibfnamefont
  {G.}~\bibnamefont {Li}}, \bibinfo {author} {\bibfnamefont {J.}~\bibnamefont
  {Shen}}, \bibinfo {author} {\bibfnamefont {Z.~Y.}\ \bibnamefont {Chen}},
  \bibinfo {author} {\bibfnamefont {Y.}~\bibnamefont {Qi}}, \bibinfo {author}
  {\bibfnamefont {Y.}~\bibnamefont {Wan}}, \bibinfo {author} {\bibfnamefont
  {W.~T.}\ \bibnamefont {Jin}}, \bibinfo {author} {\bibfnamefont
  {W.}~\bibnamefont {Li}}, \bibinfo {author} {\bibfnamefont {P.~J.}\
  \bibnamefont {Sun}},\ and\ \bibinfo {author} {\bibfnamefont {G.}~\bibnamefont
  {Su}},\ }\bibfield  {title} {\bibinfo {title} {Giant magnetocaloric effect in
  spin supersolid candidate {N}a$_2${B}a{C}o({PO}$_4$)$_2$},\ }\href@noop {}
  {\bibfield  {journal} {\bibinfo  {journal} {Nature}\ }\textbf {\bibinfo
  {volume} {625}},\ \bibinfo {pages} {270} (\bibinfo {year}
  {2024})}\BibitemShut {NoStop}%
\bibitem [{\citenamefont {Sheng}\ and\ \citenamefont
  {et~al.}(2022)}]{Sheng_PNAS_119_e2211193119}%
  \BibitemOpen
  \bibfield  {author} {\bibinfo {author} {\bibfnamefont {J.~M.}\ \bibnamefont
  {Sheng}}\ and\ \bibinfo {author} {\bibnamefont {et~al.}},\ }\bibfield
  {title} {\bibinfo {title} {Two-dimensional quantum universality in the
  spin-1/2 triangular-lattice quantum antiferromagnet
  {N}a$_2${B}a{C}o({PO}$_4$)$_2$},\ }\href@noop {} {\bibfield  {journal}
  {\bibinfo  {journal} {Proc. Natl. Acad. Sci.}\ }\textbf {\bibinfo {volume}
  {119}},\ \bibinfo {pages} {e2211193119} (\bibinfo {year} {2022})}\BibitemShut
  {NoStop}%
\bibitem [{\citenamefont {Zhong}\ \emph {et~al.}(2020)\citenamefont {Zhong},
  \citenamefont {Guo},\ and\ \citenamefont {Cava}}]{Zhong_PRM_4_084406}%
  \BibitemOpen
  \bibfield  {author} {\bibinfo {author} {\bibfnamefont {R.~D.}\ \bibnamefont
  {Zhong}}, \bibinfo {author} {\bibfnamefont {S.}~\bibnamefont {Guo}},\ and\
  \bibinfo {author} {\bibfnamefont {R.~J.}\ \bibnamefont {Cava}},\ }\bibfield
  {title} {\bibinfo {title} {Frustrated magnetism in the layered triangular
  lattice materials {K}$_2${C}o({S}e{O}$_3$)$_2$ and
  {R}b$_2${C}o({S}e{O}$_3$)$_2$},\ }\href@noop {} {\bibfield  {journal}
  {\bibinfo  {journal} {Phys. Rev. Mater.}\ }\textbf {\bibinfo {volume} {4}},\
  \bibinfo {pages} {084406} (\bibinfo {year} {2020})}\BibitemShut {NoStop}%
\bibitem [{\citenamefont {Zhu}\ \emph {et~al.}(2024)\citenamefont {Zhu},
  \citenamefont {Romerio}, \citenamefont {Steiger}, \citenamefont {Nabi},
  \citenamefont {Murai}, \citenamefont {Ohira-Kawamura}, \citenamefont
  {Povarov}, \citenamefont {Skourski}, \citenamefont {Sibille}, \citenamefont
  {Keller}, \citenamefont {Yan}, \citenamefont {Gvasaliya},\ and\ \citenamefont
  {Zheludev}}]{Zhu_PRL_133_186704}%
  \BibitemOpen
  \bibfield  {author} {\bibinfo {author} {\bibfnamefont {M.}~\bibnamefont
  {Zhu}}, \bibinfo {author} {\bibfnamefont {V.}~\bibnamefont {Romerio}},
  \bibinfo {author} {\bibfnamefont {N.}~\bibnamefont {Steiger}}, \bibinfo
  {author} {\bibfnamefont {S.~D.}\ \bibnamefont {Nabi}}, \bibinfo {author}
  {\bibfnamefont {N.}~\bibnamefont {Murai}}, \bibinfo {author} {\bibfnamefont
  {S.}~\bibnamefont {Ohira-Kawamura}}, \bibinfo {author} {\bibfnamefont
  {K.~Y.}\ \bibnamefont {Povarov}}, \bibinfo {author} {\bibfnamefont
  {Y.}~\bibnamefont {Skourski}}, \bibinfo {author} {\bibfnamefont
  {R.}~\bibnamefont {Sibille}}, \bibinfo {author} {\bibfnamefont
  {L.}~\bibnamefont {Keller}}, \bibinfo {author} {\bibfnamefont
  {Z.}~\bibnamefont {Yan}}, \bibinfo {author} {\bibfnamefont {S.}~\bibnamefont
  {Gvasaliya}},\ and\ \bibinfo {author} {\bibfnamefont {A.}~\bibnamefont
  {Zheludev}},\ }\bibfield  {title} {\bibinfo {title} {Continuum excitations in
  a spin supersolid on a triangular lattice},\ }\href@noop {} {\bibfield
  {journal} {\bibinfo  {journal} {Phys. Rev. Lett.}\ }\textbf {\bibinfo
  {volume} {133}},\ \bibinfo {pages} {186704} (\bibinfo {year}
  {2024})}\BibitemShut {NoStop}%
\bibitem [{\citenamefont {Chen}\ and\ \citenamefont
  {et~al.}(2024)}]{Chen_arxiv_2402_15869}%
  \BibitemOpen
  \bibfield  {author} {\bibinfo {author} {\bibfnamefont {T.}~\bibnamefont
  {Chen}}\ and\ \bibinfo {author} {\bibnamefont {et~al.}},\ }\bibfield  {title}
  {\bibinfo {title} {Phase diagram and spectroscopic signatures of supersolids
  in quantum ising magnet {K}$_2${C}o({S}e{O}$_3$)$_2$},\ }\href@noop {}
  {\bibfield  {journal} {\bibinfo  {journal} {arxiv:}\ }\textbf {\bibinfo
  {volume} {2402}},\ \bibinfo {pages} {15869} (\bibinfo {year}
  {2024})}\BibitemShut {NoStop}%
\bibitem [{\citenamefont {Gao}\ \emph {et~al.}(2022)\citenamefont {Gao},
  \citenamefont {Fan}, \citenamefont {Li}, \citenamefont {Yang}, \citenamefont
  {Zeng}, \citenamefont {Sheng}, \citenamefont {Zhong}, \citenamefont {Qi},
  \citenamefont {Wan},\ and\ \citenamefont {Li}}]{Gao_NPJ_7_89}%
  \BibitemOpen
  \bibfield  {author} {\bibinfo {author} {\bibfnamefont {Y.}~\bibnamefont
  {Gao}}, \bibinfo {author} {\bibfnamefont {Y.~C.}\ \bibnamefont {Fan}},
  \bibinfo {author} {\bibfnamefont {H.}~\bibnamefont {Li}}, \bibinfo {author}
  {\bibfnamefont {F.}~\bibnamefont {Yang}}, \bibinfo {author} {\bibfnamefont
  {X.~T.}\ \bibnamefont {Zeng}}, \bibinfo {author} {\bibfnamefont {X.~L.}\
  \bibnamefont {Sheng}}, \bibinfo {author} {\bibfnamefont {R.~D.}\ \bibnamefont
  {Zhong}}, \bibinfo {author} {\bibfnamefont {Y.}~\bibnamefont {Qi}}, \bibinfo
  {author} {\bibfnamefont {Y.}~\bibnamefont {Wan}},\ and\ \bibinfo {author}
  {\bibfnamefont {W.}~\bibnamefont {Li}},\ }\bibfield  {title} {\bibinfo
  {title} {Spin supersolidity in nearly ideal easy-axis triangular quantum
  antiferromagnet {N}a$_2${B}a{C}o({PO}$_4$)$_2$},\ }\href@noop {} {\bibfield
  {journal} {\bibinfo  {journal} {npj Quantum Mater.}\ }\textbf {\bibinfo
  {volume} {7}},\ \bibinfo {pages} {89} (\bibinfo {year} {2022})}\BibitemShut
  {NoStop}%
\bibitem [{\citenamefont {Wang}\ \emph {et~al.}(2001)\citenamefont {Wang},
  \citenamefont {Plackowski},\ and\ \citenamefont
  {Junod}}]{Wang_PhysicaC_355_179}%
  \BibitemOpen
  \bibfield  {author} {\bibinfo {author} {\bibfnamefont {Y.~X.}\ \bibnamefont
  {Wang}}, \bibinfo {author} {\bibfnamefont {T.}~\bibnamefont {Plackowski}},\
  and\ \bibinfo {author} {\bibfnamefont {A.}~\bibnamefont {Junod}},\ }\bibfield
   {title} {\bibinfo {title} {Specific heat in the superconducting and normal
  state (2-300 {K}, 0-16 {T}), and magnetic susceptibility of the 38 {K}
  superconductor {M}g{B}$_2$: evidence for a multicomponent gap},\ }\href@noop
  {} {\bibfield  {journal} {\bibinfo  {journal} {Physica C}\ }\textbf {\bibinfo
  {volume} {355}},\ \bibinfo {pages} {179} (\bibinfo {year}
  {2001})}\BibitemShut {NoStop}%
\bibitem [{\citenamefont {Taylor}\ \emph {et~al.}(2007)\citenamefont {Taylor},
  \citenamefont {Carrington},\ and\ \citenamefont
  {Schlueter}}]{Taylor_PRL_99_057001}%
  \BibitemOpen
  \bibfield  {author} {\bibinfo {author} {\bibfnamefont {O.~J.}\ \bibnamefont
  {Taylor}}, \bibinfo {author} {\bibfnamefont {A.}~\bibnamefont {Carrington}},\
  and\ \bibinfo {author} {\bibfnamefont {J.~A.}\ \bibnamefont {Schlueter}},\
  }\bibfield  {title} {\bibinfo {title} {Specific-heat measurements of the gap
  structure of the organic superconductors
  $\kappa$-({ET})$_2${C}u[{N}({CN})$_2$]{B}r and
  $\kappa$-({ET})$_2${C}u({NCS})$_2$},\ }\href@noop {} {\bibfield  {journal}
  {\bibinfo  {journal} {Phys. Rev. Lett.}\ }\textbf {\bibinfo {volume} {99}},\
  \bibinfo {pages} {057001} (\bibinfo {year} {2007})}\BibitemShut {NoStop}%
\bibitem [{\citenamefont {Susuki}\ \emph {et~al.}(2013)\citenamefont {Susuki},
  \citenamefont {Kurita}, \citenamefont {Tanaka}, \citenamefont {Nojiri},
  \citenamefont {Matsuo}, \citenamefont {Kindo},\ and\ \citenamefont
  {Tanaka}}]{Susuki_PRL_110_267201}%
  \BibitemOpen
  \bibfield  {author} {\bibinfo {author} {\bibfnamefont {T.}~\bibnamefont
  {Susuki}}, \bibinfo {author} {\bibfnamefont {N.}~\bibnamefont {Kurita}},
  \bibinfo {author} {\bibfnamefont {T.}~\bibnamefont {Tanaka}}, \bibinfo
  {author} {\bibfnamefont {H.}~\bibnamefont {Nojiri}}, \bibinfo {author}
  {\bibfnamefont {A.}~\bibnamefont {Matsuo}}, \bibinfo {author} {\bibfnamefont
  {K.}~\bibnamefont {Kindo}},\ and\ \bibinfo {author} {\bibfnamefont
  {H.}~\bibnamefont {Tanaka}},\ }\bibfield  {title} {\bibinfo {title}
  {Magnetization process and collective excitations in the ${S}=1/2$
  triangular-lattice {H}eisenberg antiferromagnet
  {B}a$_3${C}o{S}b$_2${O}$_9$},\ }\href@noop {} {\bibfield  {journal} {\bibinfo
   {journal} {Phys. Rev. Lett.}\ }\textbf {\bibinfo {volume} {110}},\ \bibinfo
  {pages} {267201} (\bibinfo {year} {2013})}\BibitemShut {NoStop}%
\bibitem [{\citenamefont {Xing}\ \emph {et~al.}(2019)\citenamefont {Xing},
  \citenamefont {Sanjeewa}, \citenamefont {Kim}, \citenamefont {Stewart},
  \citenamefont {Podlesnyak},\ and\ \citenamefont
  {Sefat}}]{Xing_PRB_100_220407}%
  \BibitemOpen
  \bibfield  {author} {\bibinfo {author} {\bibfnamefont {J.}~\bibnamefont
  {Xing}}, \bibinfo {author} {\bibfnamefont {L.~D.}\ \bibnamefont {Sanjeewa}},
  \bibinfo {author} {\bibfnamefont {J.}~\bibnamefont {Kim}}, \bibinfo {author}
  {\bibfnamefont {G.~R.}\ \bibnamefont {Stewart}}, \bibinfo {author}
  {\bibfnamefont {A.}~\bibnamefont {Podlesnyak}},\ and\ \bibinfo {author}
  {\bibfnamefont {A.~S.}\ \bibnamefont {Sefat}},\ }\bibfield  {title} {\bibinfo
  {title} {Field-induced magnetic transition and spin fluctuations in the
  quantum spin-liquid candidate {C}s{Y}b{S}e$_2$},\ }\href@noop {} {\bibfield
  {journal} {\bibinfo  {journal} {Phys. Rev. B}\ }\textbf {\bibinfo {volume}
  {100}},\ \bibinfo {pages} {220407(R)} (\bibinfo {year} {2019})}\BibitemShut
  {NoStop}%
\bibitem [{\citenamefont {Ono}\ \emph {et~al.}(2003)\citenamefont {Ono},
  \citenamefont {Tanaka}, \citenamefont {Katori}, \citenamefont {Ishikawa},
  \citenamefont {Mitamura},\ and\ \citenamefont {Goto}}]{Ono_PRB_67_104431}%
  \BibitemOpen
  \bibfield  {author} {\bibinfo {author} {\bibfnamefont {T.}~\bibnamefont
  {Ono}}, \bibinfo {author} {\bibfnamefont {H.}~\bibnamefont {Tanaka}},
  \bibinfo {author} {\bibfnamefont {H.~A.}\ \bibnamefont {Katori}}, \bibinfo
  {author} {\bibfnamefont {F.}~\bibnamefont {Ishikawa}}, \bibinfo {author}
  {\bibfnamefont {H.}~\bibnamefont {Mitamura}},\ and\ \bibinfo {author}
  {\bibfnamefont {T.}~\bibnamefont {Goto}},\ }\bibfield  {title} {\bibinfo
  {title} {Magnetization plateau in the frustrated quantum spin system
  {C}s$_2${C}u{B}r$_4$},\ }\href@noop {} {\bibfield  {journal} {\bibinfo
  {journal} {Phys. Rev. B}\ }\textbf {\bibinfo {volume} {67}},\ \bibinfo
  {pages} {104431} (\bibinfo {year} {2003})}\BibitemShut {NoStop}%
\bibitem [{\citenamefont {Sakhratov}\ \emph {et~al.}(2019)\citenamefont
  {Sakhratov}, \citenamefont {Prinz-Zwick}, \citenamefont {Wilson},
  \citenamefont {B  ttgen}, \citenamefont {Shapiro}, \citenamefont {Svistov},\
  and\ \citenamefont {Reyes}}]{Sakhratov_PRB_99_024419}%
  \BibitemOpen
  \bibfield  {author} {\bibinfo {author} {\bibfnamefont {Y.~A.}\ \bibnamefont
  {Sakhratov}}, \bibinfo {author} {\bibfnamefont {M.}~\bibnamefont
  {Prinz-Zwick}}, \bibinfo {author} {\bibfnamefont {D.}~\bibnamefont {Wilson}},
  \bibinfo {author} {\bibfnamefont {N.}~\bibnamefont {B  ttgen}}, \bibinfo
  {author} {\bibfnamefont {A.~Y.}\ \bibnamefont {Shapiro}}, \bibinfo {author}
  {\bibfnamefont {L.~E.}\ \bibnamefont {Svistov}},\ and\ \bibinfo {author}
  {\bibfnamefont {A.~P.}\ \bibnamefont {Reyes}},\ }\bibfield  {title} {\bibinfo
  {title} {Magnetic structure of the triangular antiferromagnet
  {R}b{F}e({M}o{O}$_4$)$_2$ weakly doped with nonmagnetic {K}$^+$ ions studied
  by {NMR}},\ }\href@noop {} {\bibfield  {journal} {\bibinfo  {journal} {Phys.
  Rev. B}\ }\textbf {\bibinfo {volume} {99}},\ \bibinfo {pages} {024419}
  (\bibinfo {year} {2019})}\BibitemShut {NoStop}%
\bibitem [{\citenamefont {Xu}\ \emph {et~al.}(2025{\natexlab{a}})\citenamefont
  {Xu}, \citenamefont {Wu}, \citenamefont {Chen}, \citenamefont {Huang},
  \citenamefont {Hu}, \citenamefont {Shi}, \citenamefont {Du}, \citenamefont
  {Li}, \citenamefont {Bian}, \citenamefont {Yu}, \citenamefont {Cui},
  \citenamefont {Zhou},\ and\ \citenamefont {Yu}}]{Xu_arxiv_2504_08570}%
  \BibitemOpen
  \bibfield  {author} {\bibinfo {author} {\bibfnamefont {X.~Y.}\ \bibnamefont
  {Xu}}, \bibinfo {author} {\bibfnamefont {Z.~L.}\ \bibnamefont {Wu}}, \bibinfo
  {author} {\bibfnamefont {Y.}~\bibnamefont {Chen}}, \bibinfo {author}
  {\bibfnamefont {Q.}~\bibnamefont {Huang}}, \bibinfo {author} {\bibfnamefont
  {Z.}~\bibnamefont {Hu}}, \bibinfo {author} {\bibfnamefont {X.~Y.}\
  \bibnamefont {Shi}}, \bibinfo {author} {\bibfnamefont {K.~F.}\ \bibnamefont
  {Du}}, \bibinfo {author} {\bibfnamefont {S.}~\bibnamefont {Li}}, \bibinfo
  {author} {\bibfnamefont {R.}~\bibnamefont {Bian}}, \bibinfo {author}
  {\bibfnamefont {R.}~\bibnamefont {Yu}}, \bibinfo {author} {\bibfnamefont
  {Y.}~\bibnamefont {Cui}}, \bibinfo {author} {\bibfnamefont {H.~D.}\
  \bibnamefont {Zhou}},\ and\ \bibinfo {author} {\bibfnamefont {W.~Q.}\
  \bibnamefont {Yu}},\ }\bibfield  {title} {\bibinfo {title} {Nmr study of
  supersolid phases in the triangular-lattice antiferromagnet
  {N}a$_2${B}a{C}o({PO}$_4$)$_2$},\ }\href@noop {} {\bibfield  {journal}
  {\bibinfo  {journal} {arxiv:}\ }\textbf {\bibinfo {volume} {2504}},\ \bibinfo
  {pages} {08570} (\bibinfo {year} {2025}{\natexlab{a}})}\BibitemShut {NoStop}%
\bibitem [{\citenamefont {Xu}\ \emph {et~al.}(2025{\natexlab{b}})\citenamefont
  {Xu}, \citenamefont {Hasik}, \citenamefont {Ponsioen},\ and\ \citenamefont
  {Nevidomskyy}}]{Xu_PRB_111_L060402}%
  \BibitemOpen
  \bibfield  {author} {\bibinfo {author} {\bibfnamefont {Y.}~\bibnamefont
  {Xu}}, \bibinfo {author} {\bibfnamefont {J.}~\bibnamefont {Hasik}}, \bibinfo
  {author} {\bibfnamefont {B.}~\bibnamefont {Ponsioen}},\ and\ \bibinfo
  {author} {\bibfnamefont {A.~H.}\ \bibnamefont {Nevidomskyy}},\ }\bibfield
  {title} {\bibinfo {title} {Simulating spin dynamics of supersolid states in a
  quantum ising magnet},\ }\href@noop {} {\bibfield  {journal} {\bibinfo
  {journal} {Phys. Rev. B}\ }\textbf {\bibinfo {volume} {111}},\ \bibinfo
  {pages} {L060402} (\bibinfo {year} {2025}{\natexlab{b}})}\BibitemShut
  {NoStop}%
\bibitem [{\citenamefont {Sheng}\ \emph {et~al.}(2025)\citenamefont {Sheng},
  \citenamefont {Wang}, \citenamefont {Jiang}, \citenamefont {Ge},
  \citenamefont {Zhao}, \citenamefont {Li}, \citenamefont {Kofu}, \citenamefont
  {Yu}, \citenamefont {Zhu}, \citenamefont {Mei}, \citenamefont {Wang},\ and\
  \citenamefont {Wu}}]{Sheng_Innovation_6_100769}%
  \BibitemOpen
  \bibfield  {author} {\bibinfo {author} {\bibfnamefont {J.~M.}\ \bibnamefont
  {Sheng}}, \bibinfo {author} {\bibfnamefont {L.}~\bibnamefont {Wang}},
  \bibinfo {author} {\bibfnamefont {W.~R.}\ \bibnamefont {Jiang}}, \bibinfo
  {author} {\bibfnamefont {H.}~\bibnamefont {Ge}}, \bibinfo {author}
  {\bibfnamefont {N.}~\bibnamefont {Zhao}}, \bibinfo {author} {\bibfnamefont
  {T.~T.}\ \bibnamefont {Li}}, \bibinfo {author} {\bibfnamefont
  {M.}~\bibnamefont {Kofu}}, \bibinfo {author} {\bibfnamefont {D.~H.}\
  \bibnamefont {Yu}}, \bibinfo {author} {\bibfnamefont {W.}~\bibnamefont
  {Zhu}}, \bibinfo {author} {\bibfnamefont {J.~W.}\ \bibnamefont {Mei}},
  \bibinfo {author} {\bibfnamefont {Z.~T.}\ \bibnamefont {Wang}},\ and\
  \bibinfo {author} {\bibfnamefont {L.~S.}\ \bibnamefont {Wu}},\ }\bibfield
  {title} {\bibinfo {title} {Continuum of spin excitations in an ordered
  magnet},\ }\href@noop {} {\bibfield  {journal} {\bibinfo  {journal} {The
  Innovation}\ }\textbf {\bibinfo {volume} {6}},\ \bibinfo {pages} {100769}
  (\bibinfo {year} {2025})}\BibitemShut {NoStop}%
\bibitem [{\citenamefont {Jia}\ \emph {et~al.}(2024)\citenamefont {Jia},
  \citenamefont {Ma}, \citenamefont {Wang},\ and\ \citenamefont
  {Chen}}]{Jia_PRR_6_033031}%
  \BibitemOpen
  \bibfield  {author} {\bibinfo {author} {\bibfnamefont {H.~C.}\ \bibnamefont
  {Jia}}, \bibinfo {author} {\bibfnamefont {B.~W.}\ \bibnamefont {Ma}},
  \bibinfo {author} {\bibfnamefont {Z.~D.}\ \bibnamefont {Wang}},\ and\
  \bibinfo {author} {\bibfnamefont {G.}~\bibnamefont {Chen}},\ }\bibfield
  {title} {\bibinfo {title} {Quantum spin supersolid as a precursory dirac spin
  liquid in a triangular lattice antiferromagnet},\ }\href@noop {} {\bibfield
  {journal} {\bibinfo  {journal} {Phys. Rev. Research}\ }\textbf {\bibinfo
  {volume} {6}},\ \bibinfo {pages} {033031} (\bibinfo {year}
  {2024})}\BibitemShut {NoStop}%
\end{thebibliography}

\end{document}